\documentclass[12pt]{article}

\usepackage{newtxtext,newtxmath}

\usepackage{graphicx}
\usepackage{subcaption}
\usepackage[letterpaper,margin=1in]{geometry}

\renewenvironment{abstract}
	{\quotation}
	{\endquotation}

\date{}

\makeatletter
\renewcommand{\fnum@figure}{\textbf{Figure \thefigure}}
\renewcommand{\fnum@table}{\textbf{Table \thetable}}
\makeatother

\usepackage{scicite}

\usepackage{url}

\def\scititle{Linear Superposition Effect at Sources and in Waves}
\title{\bfseries \boldmath \scititle}

\author{
	Bingli JIAO\small$^{1,2}$\\
	\small$^{1}$School of Electronics, Peking University, Beijing 100871, China.\\	\small$^\ast$jiaobl@pku.edu.cn\\
    \small$^{2}$Department of Electrical and Computer Engineering, Princeton University, Princeton 08544, America.\\	
    \small$^\dagger$bjiao@princeton.edu\\
}

\begin{document} 

\maketitle

\begin{abstract} \bfseries \boldmath

The superposition law (SL) sums the components of electromagnetic (EM) waves at each spatial point when these waves meet in space. In contrast, the energy conservation law requires energy to be summed in the quadratic form of the EM fields.  The mathematical discrepancy of the two laws can lead to different physical results.  Specifically, when two co-phase radiation dipoles are placed in close proximity, their radiation waves undergo a co-phase interference throughout space, therefore causing a net increase in  wave's power globally.  In the  exploration of this, we find that the SL applies not only to waves, but also to the radiation dipoles.  By defining the superposed dipole conceptually, we describe the effective radiation power that is twice the power-sum of the two waves, providing a comprehensive understanding of the SL, which is supported by the results of the previous theoretical and experimental studies.

\end{abstract}

\section*{Effect of Linear Superposition }

\label{Process of Reserving Energy Conservation}

\noindent
The Superposition Law (SL) governs the wave-behavior when two or more waves meet in space, and its linear additivity has been confirmed by experiments. In comparison, the Energy Conservation Law (ECL) can be regarded as a law of nonlinear addition — more precisely, one with a squared superposition property — and to some extent it is an empirical law. 

When the SL is applied to two electromagnetic (EM) waves of the same frequency, the resultant field can be expressed as   
\begin{equation}
	\begin{split}\label{S-1}
		\overrightarrow{E}(x,y,z,\omega t) = \overrightarrow{E}_1(x,y,z,\omega t) +\overrightarrow{E}_2(x,y,z,\omega t) 
	\end{split}
 \end{equation}   
where $\omega$ is the angular frequency, and $x,y,z$ are the Cartesian coordinates.  The vectors $\overrightarrow{E}_1(x,y,z)$ and   $\overrightarrow{E}_2(x,y,z)$ represent the electric field strengths of waves 1 and 2, and 
$\overrightarrow{E}(x,y,z)$ that of their superposition, respectively.  

Equation \eqref{S-1} relates to the energy metric by taking the square operation and the time average as
\begin{equation}\label{S-E}
	<E^2(x,y,z)> = <{E}_1^2(x,y,z)> + <{E}_2^2(x,y,z)> + <{\Gamma}(\phi)> 
\end{equation}  
with  
\begin{equation}\label{Inter}
	\Gamma(\phi) = 2 E_1(x,y,z) E_2(x,y,z)\text{cos} [\phi (x,y,z)]
\end{equation} 
where $\langle \cdot\rangle$ represents the operator of time averaging, $\Gamma(\phi)$ the interference term and $\phi$ the phase difference between the two radiation waves. 

However, it is found that the energy conservation is a special case of \eqref{S-E}, expressed as  
\begin{equation}\label{ECL-G}
	<{E}^2(x,y,z)> = <{E}_1^2(x,y,z)> + <{E}_2^2(x,y,z)>,
\end{equation}  
which definitely requires the nullification of the interference, i.e., $\Gamma(\phi) \equiv 0$ in \eqref{Inter}.  However, since both theoretical analyses and experimental observations substantiate the cases of non-zero values of the interference term, the condition $\Gamma(\phi) \equiv 0$ cannot, in general, be sustained.  Thus, the validity of the SL stands in fundamental conflict with the ECL at the local level, often expressed in terms of power densities in the Poynting theorem.    

In order to preserve the ECL, researchers attempt to eliminate the local energy abnormalities in \eqref{S-E} by taking the spatial average, which is expressed as
\begin{equation}\label{S-Av-E}
	\mathfrak{A} \{<E^2>\} = \mathfrak{A} \{ <E^2_1>\} + \mathfrak{A} \{<E_2^2>\} + \mathfrak{A} \{ <\Gamma(\phi)>\} ,  
\end{equation}  
with an expectation that   
\begin{equation}\label{S-Av-In}
\mathfrak{A} \{ \Gamma(\phi)\} \equiv 0
\end{equation}
where $\mathfrak{A}\{\cdot\}$ represents an operator that calculates the spatial average value over entire space.

Unfortunately, \eqref{S-Av-In} fails in the model studied by Levine in 1980 [1], where the co-phase interference, i.e., $\phi \approx 0$, occurs through space, resulting in 
\begin{equation}\label{S-Av-In-C}
	\mathfrak{A} \{ \Gamma(\phi)\} \approx [\mathfrak{A} \{<E^2_1>
	\} + \mathfrak{A} \{<E^2_2> \}]	
\end{equation}
with assumption of $ \mathfrak{A} \{<E^2_1>\} = \mathfrak{A} \{<E^2_2> \}$, instead of being zero. 

As a result, the energy is doubled at the stage of wave's description, as expressed by  
\begin{equation}\label{two}
	\mathfrak{A} \{<E^2>\} = 2 [\mathfrak{A} \{<E^2_1>
	\} + \mathfrak{A} \{<E^2_2> \}]	  \ \ \   
\end{equation}   
where $\mathfrak{A} \{<E^2>\}$ is the total radiation power and $``2"$ on the right side of the equality is the energy-doubling factor.  Actually, the concept of this energy-doubling has been accepted there [1-4].

It is worthy of mention that the abnormal phenomenon reflected in \eqref{two} was incorrectly attributed to the EM coupling between the sources [1].  This mistake was clarified by researchers who arranged the two sources in a spatially symmetric structure, whereby no matter what EM coupling occurs, the symmetry allows the two sources to radiate in a co-phase manner.  In this case, the energy-doubling can still occur at the wave stage.  Taking one step further logically, if the energy of the waves is doubled, then the energy of the sources must also be doubled [2]. This is the issue that we will focus on through further analysis.    

We may infer that the energy behavior above is a property of wave superposition including those with the probability wave, as were observed not only in classical electromagnetic fields [5,6], but also in quantum mechanics [7–9].

\section*{Effective Dipole}
To be specific, we design a symmetric radiation system in free space and represent the two identical dipole antennas as vectors without loss of generality, as illustrated by the snapshot in Fig. 1(a)  , to radiate the same power in a co-phase manner.
$\mathfrak{G}$ denotes a spherical surface that restricts our study to the far-field regime, with the conditions given by 
\begin{equation}\label{condition}
	\ \ \ \  \ \ |l + 2d| << \lambda \ \ \text{and} \  \ \ \lambda << r \ \ \ \ \ \ \ \text{for}  \ \ \ \ \   l,d \ne 0 \, \ 
\end{equation} 
to guarantee the occurrence of co-phase interference of the waves throughout space, where $l$ is the distance between the two charges in each dipole, $d$ is the distance between the two dipoles, $\lambda$ is the wavelength and $r$ is the radius in spherical coordinates.  

To begin, we examine one radiating dipole and its radiation field while temporarily neglecting the presence of the other dipole. The electric field can be written as [9]     
\begin{equation}\label{S-E-2}
	<\overrightarrow{E}_i> = j\frac{\omega \beta \eta (lq) }{ 4\pi r_i} e^{-j \beta r_i}sin(\theta_i)\overrightarrow{e}_{\theta_1}  \ \ \ \text{for} \ \ i=1 \ \ \text{or}\ \ 2
\end{equation} 
where $\overrightarrow{E}_i$ denotes the electric field strength generated by dipole $i$, $\omega$ is the angular frequency, $\beta= 2\pi/ \lambda$ is the wave number, $\eta_0 = \sqrt{(\mu_0/\epsilon_0)}$ is the intrinsic impedance, $r_i$ is distance from dipole $i$ to observation point on the spherical surface and $\overrightarrow{e}_{\theta_i}$ is the direction of the  electric field, as shown in Fig. 1(a).  

A key conclusion drawn from \eqref{S-E-2} is the fact that the power flow from dipole $i$ vanishes along the $z$-axis (since $\theta_i = 0$ or $ \pi$), which implies that the EM coupling between the two dipoles is, in principle, zero.  To confirm this, refer to [2], which shows the experimental result that the EM coupling can be negligible [2].  In this regard, the individual dipole radiates as if it were alone in performing power, given by [9]  

\begin{equation}\label{Single-Power}
	P_i = \frac{\pi \omega^2}{3 \lambda^2} \eta (lq)^2
\end{equation}
where $P_i$ is the radiation power of dipole $i$.   

We note that ``${lq}$" in both Eqs. \eqref{S-E-2} and \eqref{Single-Power} represents the physical metric of the radiation dipole to be considered later. 

In the far-field regime, we work with the approximations of  $r_1 \approx r_2 \approx r$ and $\overrightarrow{e}_{\theta_1} \approx \overrightarrow{e}_{\theta_2} \approx  \overrightarrow{e}_{\theta} $, under which the superposition of the two electric fields expressed in \eqref{S-E-2} yields        
\begin{equation}\label{S-E-S}
	<\overrightarrow{E}_s> \approx j\frac{\omega \beta \eta (2lq) }{ 4\pi r} sin (\theta) e^{-j \beta r} \ \overrightarrow{e}_{\theta},  
\end{equation} 
where $<\overrightarrow{E}_s>$ is the resultant field.

By comparing \eqref{S-E-S} with \eqref{S-E-2}, we find that the radiation dipole changes from ``$lq$" to ``$2lq$", demonstrating the effect of linear superposition.  By referring to ``$2lq$" as an effective dipole, we find further that the total radiation power is doubled by considering Eq. \eqref{Single-Power}  
\begin{equation}\label{Single-P}
	P_s = \frac{\pi \omega^2}{3 \lambda^2} \eta (2lq)^2 = 4P_i = 2(P_1 + P_2)
\end{equation}
where $P_s$ is defined as the effective radiation power (ERP) of the two dipoles, doubling the sum-power of the individual waves.
   
Since there is no EM coupling between the two dipoles, the radiation power of each dipole remains unchanged, as shown in \eqref{Single-Power}. Therefore, we turn to accept the fact that the ERP results from the superposition of the dipoles.  To complete the description of this, we denote dipole $i$ by    $\overrightarrow{\mathfrak{D}}_i = lq\overrightarrow{k}$ and explain the dipole-superposition in \eqref{S-E-S} by 
\begin{equation}\label{S-S}
	\overrightarrow{\mathfrak{D}}_s = \overrightarrow{\mathfrak{D}}_1 + \overrightarrow{{\mathfrak{D}}_2} = 2lq \overrightarrow{k}
\end{equation} 
where $\overrightarrow{\mathfrak{D}}_1$,  $\overrightarrow{\mathfrak{D}}_2$  and $\overrightarrow{\mathfrak{D}}_s$ are the radiation dipoles 1 and 2 and the effective dipole, respectively.  The process of superposition is shown in Fig. 1 (a) and (b), where it can be found that the effective dipole $\overrightarrow{\mathfrak{D}_s}$ is independent of the distance $d$ between the two dipoles under the assumption of Eq.\eqref{condition}. 

It is noted that the above derivations are consistent with Noether’s theorem for measuring the energy in the time domain.

\begin{figure}[!t]
	\centering
	\begin{subfigure}[b]{0.45\textwidth}
		\centering
		\includegraphics[width=\linewidth]{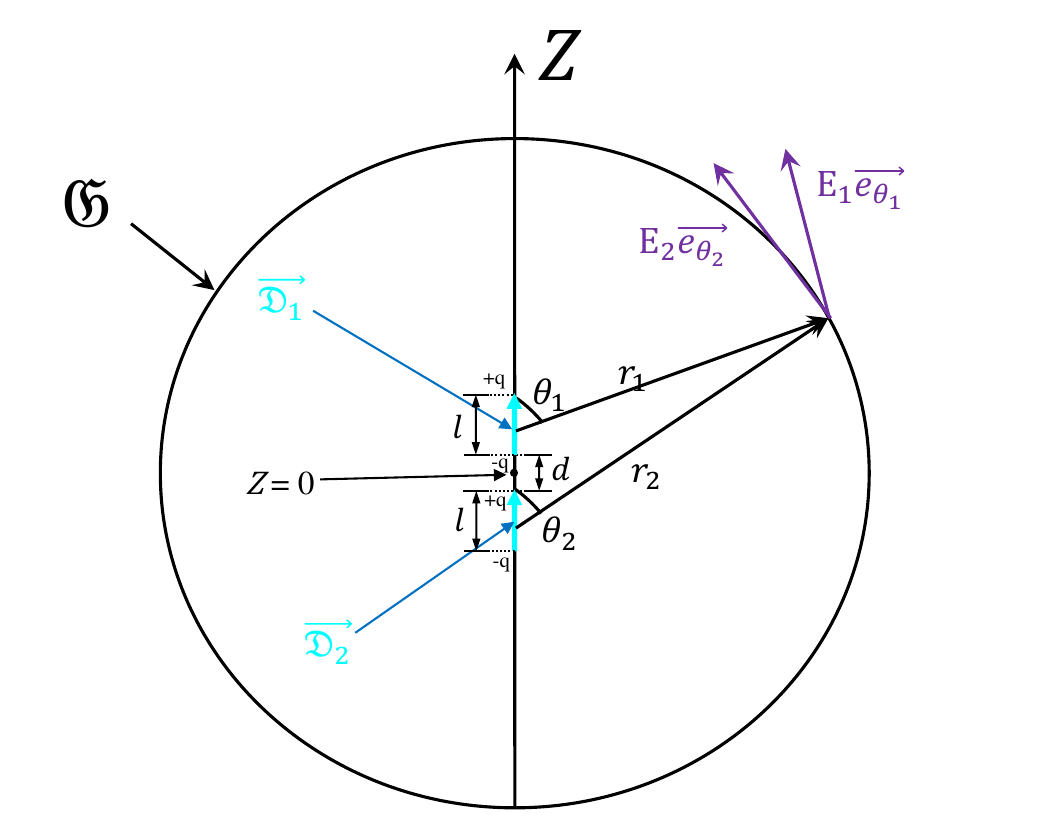}
		\caption{The vectorized dipoles and the electric fields.}
		\label{fig_radiation_1} 
	\end{subfigure}
	\hfill
	\begin{subfigure}[b]{0.45\textwidth}
		\centering
		\includegraphics[width=\linewidth]{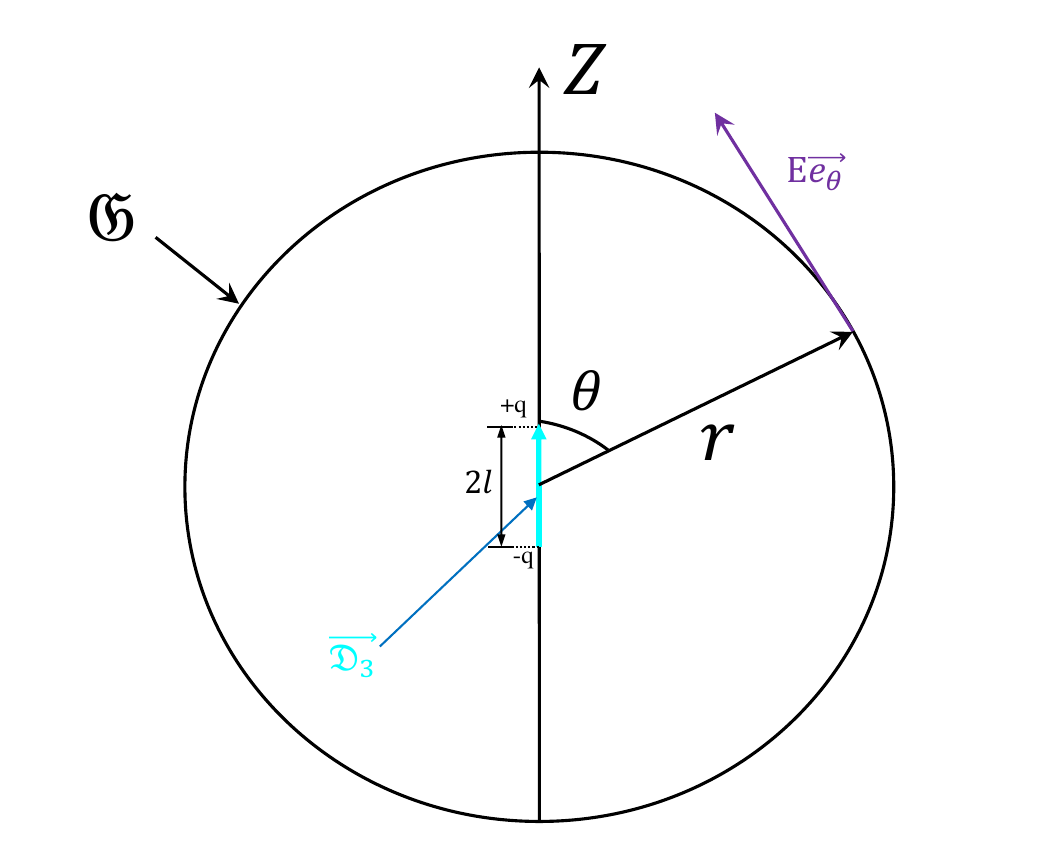}
		\caption{Superposed dipole and its electric field.}
		\label{fig_radiation_1} 
	\end{subfigure}
	
	\caption{The diagram showing the process of the dipole superposition.}
	\label{fig_radiation_1} 
\end{figure}

\section*{Result and Discussion}
To substantiate the energy-doubling result, we provide the following discussions. 
\begin{itemize}
    \item Wave superposition
    
    The proposed system enables the interference of the two waves in a co-phase interference at every point in space, thereby doubling radiation power globally. 
    
    \item Source superposition
    
    The SL is validated for its application at the two radiation dipoles, identifying the fundamental reason for the two-fold increase in radiation power. Hence, we create the concept of ERP to describe the energy effect of the linear superposition.
   
	\item Domination of the LS 
	
	Although the ECL requires that the total power equals power-sum of the two dipoles, the SL dominates and results in the energy-doubling both at the sources and in the waves.

\end{itemize}

Finally, the energy-doubling result is confirmed by a 1.58-fold increase in energy observed in the previous experiment with the same system design [3].


\end{document}